\documentclass[aps,prb,twocolumn,groupedaddress,showpacs]{revtex4-1}
\usepackage{amsmath}
\usepackage{graphicx}
\usepackage{epstopdf}
\usepackage{bm}
\usepackage{amssymb}
\usepackage{subfigure}
\usepackage[breaklinks=true,
colorlinks=true,
linkcolor=blue,
citecolor=blue,
urlcolor=blue]{hyperref}

\begin{document}

 \title{Enhanced strong-coupling stimulated Brillouin amplification assisted by Raman amplification}

  \author{Y. Chen$^{1,2}$, C. Y. Zheng$^{3,4,5}$, Z. J. Liu$^{3,4}$, L. H. Cao$^{3,4,5}$, and C. Z. Xiao$^{2,5}$}
  \email{xiaocz@hnu.edu.cn}
  \affiliation{$^1$ School of Electrical and Information Engineering, Anhui University of Science and Technology, Huainan,Anhui 232001, China}
  \affiliation{$^2$Key Laboratory for Micro-/Nano-Optoelectronic Devices of Ministry of Education, School of Physics and Electronics, Hunan University, Changsha, 410082, China}
  \affiliation{$^3$Institute of Applied Physics and Computational Mathematics, Beijing, 100094, China}
  \affiliation{$^4$\mbox{HEDPS, Center for Applied Physics and Technology, Peking University, Beijing 100871, China}}
  \affiliation{$^5$\mbox{Collaborative Innovation Center of IFSA (CICIFSA), Shanghai Jiao Tong University, Shanghai 200240,China}}

  \date{\today}

  \begin{abstract}

   Higher intensity of strong-coupling stimulated Brillouin scattering (SC-SBS) amplification is achieved by supplementary Raman amplification. In the new scheme, a Raman pump laser first amplifies the seed pulse in the homogeneous plasma, then  a SC-SBS pump laser continues the amplification in the inhomogeneous plasma in order to suppress the spontaneous instability of pump lasers. The intensity of seed laser gets higher  and the duration of seed laser gets shorter than that in the pure SC-SBS scheme with the same incident energy, while the energy conversion efficiency is not significantly reduced. We also found that the SC-SBS amplification is seeded by the leading pulse of Raman amplification.  The results obtained from envelope coupling equations, Vlasov simulations and two-dimensional particle-in-cell(PIC) simulations agree with each other. This scheme offers a possible way to improve the SC-SBS amplification in experiments.

  \end{abstract}


  \maketitle

\section{Introduction}\label{introduction}
Laser amplification by laser plasma instabilities is a promising and powerful way to produce high intensity laser pulse\cite{malkin1,malkin2}. Using plasma as the gain medium avoids the damage of optical grating in chirped pulse amplification(CPA) technique\cite{cpa}. Two kinds of laser plasma instabilities are commonly applied to plasma based amplification: stimulated Raman scattering (SRS)\cite{2dSRS,tsi,SRS1,SRS2,SRS3,SRS4,SRS5,SRS6,SRS7,SRS8,SRS9,SRS10,SRS11,SRS12,SRS13,SRS14,SRS15,SRS16,SRS17,rand,Riconda2d} and strong-coupling stimulated Brillouin scattering (SC-SBS)\cite{SBS1,SBS2,chen,SBS3,SBS4,SBS5,SBS6,SBS13,SBS7,SBS8,SBS9,SBS11,SBS12,Amir,Riconda,leh1,leh2,shou}. A seed laser pulse couples with a counter-propagating pump laser via a plasma wave $i.e.$ Langmuir wave or ion acoustic wave\cite{kruer,Nicholson}, then the energy of pump laser will be transferred to seed laser pulse leading to a rapid increase of seed laser intensity.

Linear and nonlinear stage exist in both Raman amplification and SC-SBS amplification. At the linear stage, the amplitude of pump laser can be considered as a constant, so we only discuss the exponential growth of seed laser and plasma waves; at the  nonlinear stage, linear growth of seed laser is observed because of the pump depletion, and the energy flows from the tail of seed laser back to pump laser, which results in the $\pi$ pulses. This stage is also called the self-similar stage\cite{SRS13,SBS6,SBS8}. The duration of leading pulse would be reduced in the nonlinear stage of Raman amplification, because the peak of seed laser is superluminal. Early works also observed this phenomenon in Raman amplification\cite{SRS5,SRS7,compress1,compress2,compress3,compress4}.  However, in the nonlinear stage of SC-SBS amplification, the reduction of seed laser duration is slower than that in Raman amplification \cite{SBS6,SBS8,SBS9,leh1,SBS10}, the pulse duration is usually hundreds of femtoseconds\cite{SBS5}. Amiranoff $et.al.$ reported that the duration of seed laser is related to the ratio of the intensities of seed laser and pump laser, so, using a seed laser with higher intensity may reduce the duration of seed laser in SC-SBS amplification.

In this paper, we propose a scheme that by adding a short-distance Raman amplification ahead of the SC-SBS amplification, the pulse width of seed laser will be substantially reduced and the maximum intensity of seed laser is increased markedly. Firstly, the  coupled envelope equations for SRS and SC-SBS are constructed, which contains five waves: SRS pump laser, SC-SBS pump laser, Langmuir wave, ion acoustic wave and seed laser pulse. The maximum intensity of amplified seed laser is increased about $36\%$ after using SRS amplification. The SC-SBS amplification is seeded by the $\pi$ pulse of SRS, because the structure of seed laser after SRS amplification is similar to that in SC-SBS amplification.  Then we discuss this phenomenon by phase analysis and find that higher initial intensity of seed laser will cause shorter energy transfer region.

Next, for a more detailed discussion of the effects of SRS amplification, one dimensional and fully kinetic Vlasov-Maxwell code (Vlama)\cite{liu} is used. Higher intensity of seed laser is obtained, which agrees well with our five-wave amplification model. We also found that the optimal proportion of SRS, $i.e.$ $\rm{L}_{\rm{SRS}}/\rm{L}_{\rm{total}}=0.25$, where $\rm{L}_{\rm{SRS}}$ is the SRS amplification length, $\rm{L}_{\rm{total}}$ is the total amplification length. Under this condition, the maximum intensity of seed laser is increased about $69\%$, and the energy transfer efficiency only decreases $3\%$, compared with pure SC-SBS amplification. It should be noticed that the total input energy of pump laser, which include the energy of SRS pump laser and SC-SBS pump laser, is the same with the pure SC-SBS case. At last, a two-dimensional particle-in-cell(PIC) code is used to investigate the influence of multidimensional effects. Similar with one dimensional cases, the intensity of amplified seed laser also increases in the new scheme. However, the filamentation of seed laser is observed in two dimensional simulations which may be harmful for the amplification.

This paper is structured in the following ways. Firstly, in Sec.~\ref{theoretical model}, we describe the envelope coupling equations of SRS and SC-SBS amplifications. Secondly, the five-wave amplification model is numerically solved to study the new scheme in Sec.~\ref{numerical model}. Thirdly,  Vlasov simulations and PIC simulations to verify our five-wave simulation results in Sec.~\ref{Simulation model} and Sec.~\ref{PIC model}. At last, the conclusion  and discussion about the new scheme and the experimental guidance are shown in  Sec.~\ref{conclusion}.

\section{Theoretical analysis of five-wave amplification model}\label{theoretical model}


\begin{figure}[htbp]
    \begin{center}
      \includegraphics[width=0.45\textwidth,clip,angle=0]{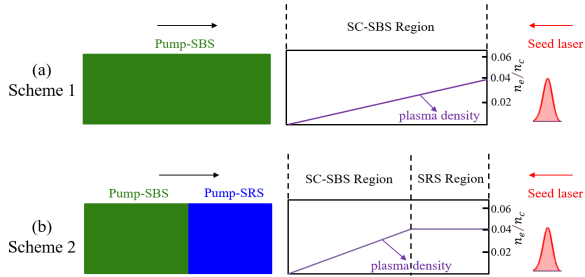}\vspace{-10pt}
      \caption{\label{scheme} (a) The pure SC-SBS amplification. (b) The five-wave amplification scheme, the plasma region is divided into SC-SBS region and SRS region, in front of SC-SBS pump laser, the SRS pump laser is added.}
    \end{center}
  \end{figure}

\begin{figure*}
    \begin{minipage}[t]{0.487\linewidth}
    \centering
    \includegraphics[width=3.65in]{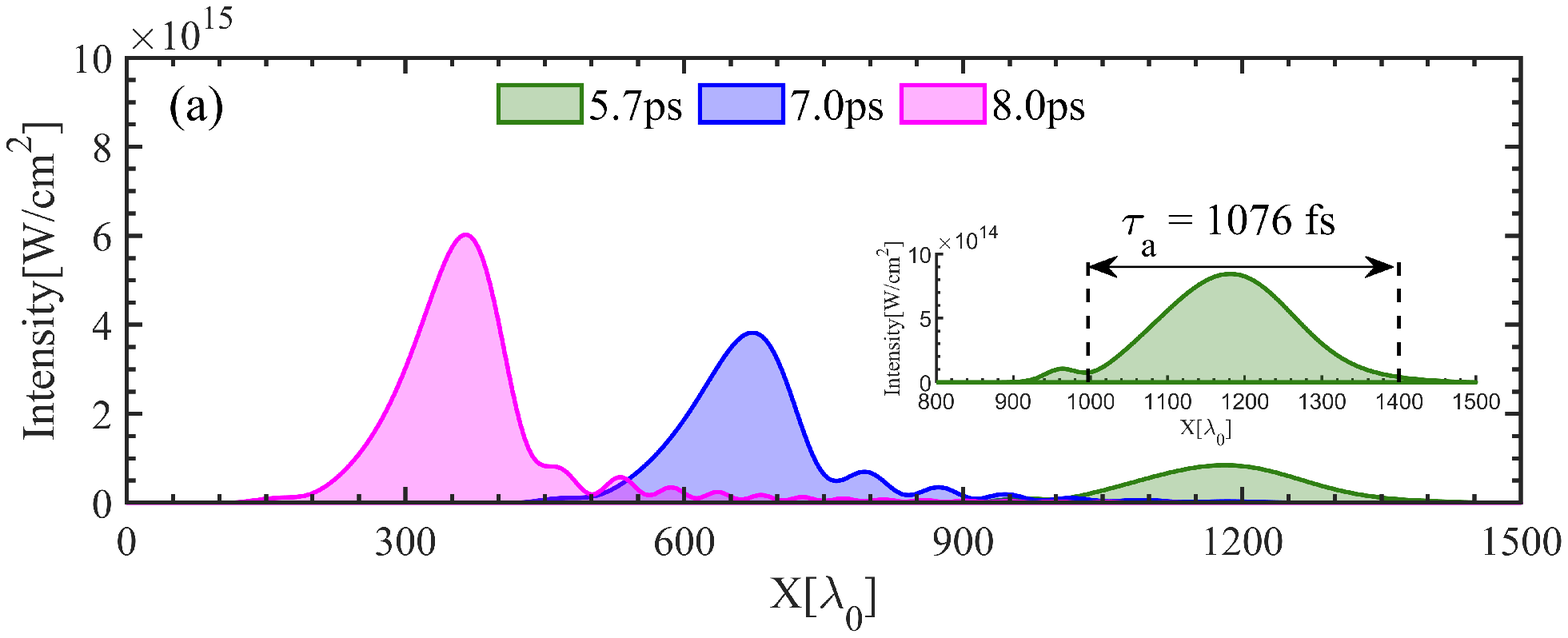}\vspace{-20pt}
    \end{minipage}%
    \begin{minipage}[t]{0.487\linewidth}
    \centering
    \includegraphics[width=3.65in]{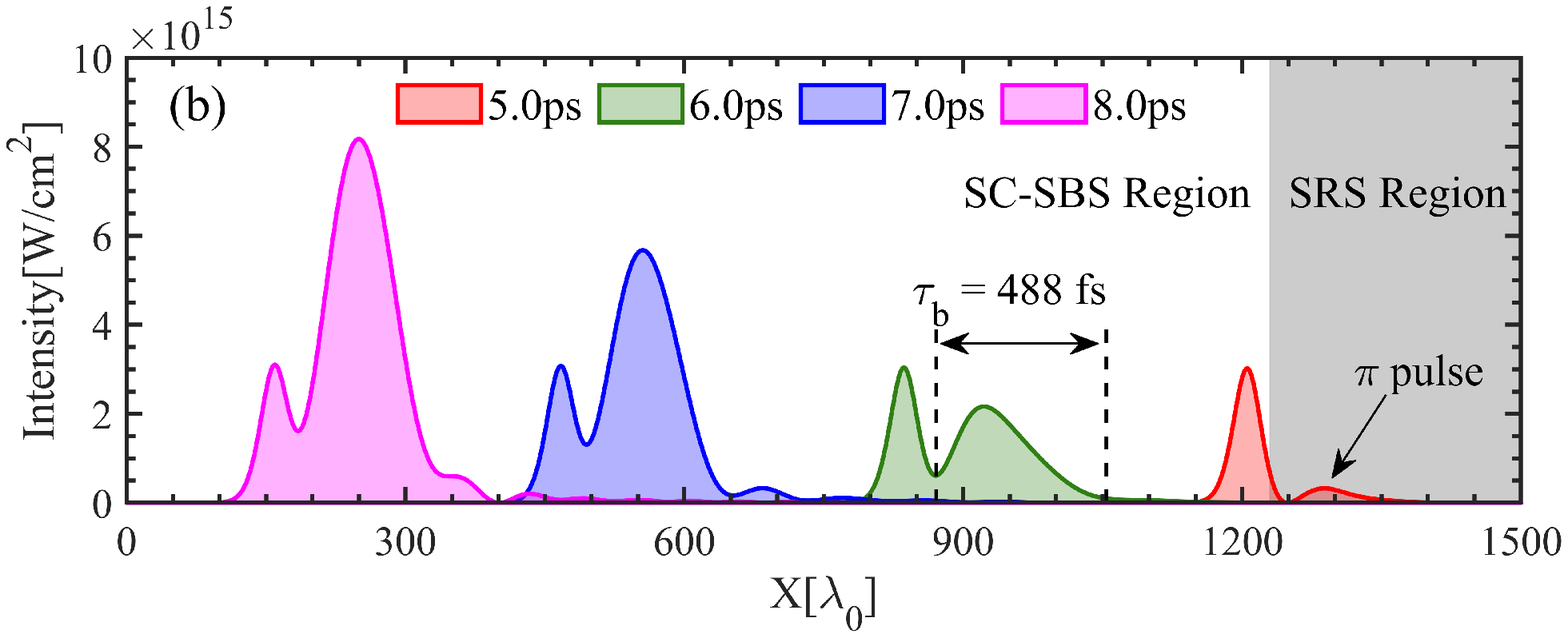}\vspace{-20pt}
    \end{minipage}

    \caption{\label{two_case_2} Numerical solution of Eq.~(\ref{wkb1}), (a) is the amplification of Scheme 1, inset figure is the seed laser at $t = 5.7$ ps (b) is the amplification of Scheme $2$ with $\rm{L}_{\rm{SRS}}/\rm{L}_{\rm{total}}=0.18$.}
\end{figure*}

Stimulated Raman scattering and strong-coupling stimulated Brillouin scattering are the commonly used laser plasma instabilities in pulse amplification. Pulse amplification by SRS has advantage;  the intensity growth of seed laser in SRS amplification is faster than that in SC-SBS amplification at the early stage, because SRS usually has a higher growth rate; However, SRS amplification also has its disadvantages; (i) the energy transfer efficiency is low, because nearly half of the pump laser's energy flows to Langmuir waves; (ii) the $\pi$ pulses appear in the nonlinear stage, and they will absorb a part of pump laser's energy.

Similarly, SC-SBS amplification has its advantages; (i) the energy transfer efficiency is higher. Most of the pump laser's energy flows to the seed laser, because the frequency of ion acoustic waves is far less than the frequency of pump laser; (ii) the amplitude of $\pi$ pulse is lower than that in Raman amplification. The disadvantage of SC-SBS amplification is the lower growth rate than SRS,  so it will take a longer time to enter the exponential growth stage\cite{Amir}.

 A natural idea is to combine the advantages of SRS and SC-SBS to obtain a better laser amplification. In this paper, we propose a scheme of SC-SBS amplification assisted by SRS amplification. As shown in Fig.~\ref{scheme}, Scheme $1$ is the normal SC-SBS amplification in an  inhomogeneous plasma. Green rectangle is the pump laser of SC-SBS, which injects from left, and Gaussian seed laser enters the plasma from the right. In the new scheme, a part of SC-SBS amplification is replaced by SRS amplification in the homogeneous plasma, correspondingly, there exists the SRS pump laser in front of SC-SBS pump laser. The duration of SRS pump laser is obtained by $\tau_{\rm{SRS}} \approx 2\rm{L}_{\rm{SRS}}/c$, where $\rm{L}_{\rm{SRS}}$ is the length of SRS region and $c$ is the light speed in vacuum.

 It should be noticed that the SC-SBS amplification in nonuniform density has three advantages: (i) it can mitigate the influence of spontaneous SRS of pump laser; (ii) the triangular density profile  adds an effective chirp on pump laser, which  is favorable to amplification\cite{SBS10}; (iii) the amplitude of $\pi$-pulse in triangular density profile is lower than that in uniform density profile, and the energy exchange mainly occurs at the first peak of seed laser\cite{Amir,SBS10}.

In order to describe Scheme $2$, we construct five-wave envelope coupling equations including both SRS and SC-SBS\cite{kruer,Nicholson},

\begin{equation} \label{wkb1}
    \begin{split}
    &(\partial_{t}+\nu_{00}+V_{00}\partial_{x})a_{00}=-\frac{i}{4} a_{1}\delta n_{epw},\\
    &(\partial_{t}+\nu_{01}+V_{01}\partial_{x})a_{01}=-\frac{i\omega_{00}}{4\omega_{1}} a_{1}\delta n_{iaw},\\
    &(\partial_{t}+\nu_{1}+V_{1}\partial_{x})a_{1}=-\frac{i\omega_{00}}{4\omega_{1}}(\delta n_{epw}^{*}a_{00}+\delta n_{iaw}^{*}a_{01} ),\\
    &(\partial_{t}+\nu_{2}+V_{2}\partial_{x})\delta n_{epw}=-\frac{4i\Gamma_{SRS}^{2}\omega_{1}c^{2}}{\omega_{00}^{3}v_{00}^{2}}a_{00}a_{1}^{*},\\
    &(\partial_{t}^{2}+V_{3}^{2}\partial_{x}^{2})\delta n_{iaw}=-\frac{Z\beta n_{e}(x)k_{iaw}^{2}c^{2}}{2 n_{c}\omega_{00}^{2}}a_{01}a_{1}^{*},
    \end{split}
   \end{equation}
   where $a_{00}$, $a_{01}$ and $a_{1}$ are slowly varying  amplitudes of SRS pump wave, SC-SBS pump wave and seed wave, respectively. $\delta n_{epw}$ and $\delta n_{iaw}$ are the electron density perturbations of Langmuir wave and ion acoustic wave, respectively, which are normalized to $n_{c}$, where $n_{c}$ is the critical density of SRS pump laser. When $a_{00}$ equals to $0$, Eq.~(\ref{wkb1}) becomes to the envelope coupling equations for Scheme 1.

   In Scheme $2$, SRS and SC-SBS share one seed laser, so waves in Eq.~(\ref{wkb1}) should  satisfy phase matching conditions,
   \begin{equation} \label{phase}
   \begin{split}
     &\omega_{00} = \omega_{1} + \omega_{epw},\\
     &\omega_{01} = \omega_{1} + \omega_{iaw},\\
     &|k_{epw}| = |k_{00}|+|k_{1}|,\\
     &|k_{iaw}| = |k_{01}|+|k_{1}|,\\
    \end{split}
   \end{equation}
    where $\omega_{00}$, $\omega_{01}$, $\omega_{1}$ , $\omega_{epw}$ and $\omega_{iaw}$ are the frequency of five waves, and $k_{00}$,$k_{01}$,$k_{1}$,$k_{epw}$ and $k_{iaw}$ are the corresponding wavenumbers. In practice, $\omega_{01}$ is usually equals to $\omega_{1}$, because $\omega_{iaw} \ll \omega_{1}$, thus, $|k_{iaw}| = 2|k_{1}|$. $V_{00}$, $V_{01}$, $V_{1}$, $V_{2}$ and $V_{3}$ are the group velocity of five waves. and $V_{1}=-V_{01}$. $\nu_{00}$, $\nu_{01}$, $\nu_{1}$ and $\nu_{2}$ are the corresponding damping rates, for simplicity, the collision dampings of lasers are neglected, and $\nu_{2}$ is the Landau damping of Langmuir wave. $Z$ is the charge state of ions, $\beta = m_{e}/m_{i}$ is the mass ratio of electrons and ions, $n_{e}(x)$ is the plasma density.

    As we know, weak coupling SBS will transit to SC-SBS because of higher intensity of pump laser, and the threshold of SC-SBS has been studied by the early work\cite{forslund},
    \begin{equation} \label{SC-SBS_threshold}
           (v_{01}/c)^{2}>4k_{01}V_{3}\omega_{1}v_{e}^{2}/(\omega_{pe}c)^{2},
    \end{equation} where $v_{01}$ is the electron oscillation velocity in SC-SBS pump laser, $\omega_{pe}$ is the plasma frequency. The density perturbation of ion acoustic wave oscillates rapidly over an acoustic period, thus we retain the second derivative term in the equation of ion acoustic wave. The maximum growth rate of SC-SBS is related to the amplitude of pump laser,
   \begin{equation} \label{SC-SBS_growth}
       \Gamma_{\rm{SC-SBS}}=\frac{\sqrt{3}}{2} \left( \frac{k_{01}^{2}v_{01}^{2} \omega_{pi}^{2}}{2\omega_{1}} \right)^{1/3},
   \end{equation}
    where $\omega_{pi}$ is the plasma ion frequency. Besides, the growth rate of SRS in Scheme 2 is\cite{Nicholson}
   \begin{equation} \label{SRS_growth}
       \Gamma_{\rm{SRS}}=\frac{k_{epw}v_{00}}{4}\left[ \frac{\omega_{pe}^{2}}{\omega_{epw}\left(\omega_{00}-\omega_{epw}\right)}\right]^{1/2},
   \end{equation}
    where $v_{00}$ is the electron oscillation velocity in SRS pump laser.

   In Scheme $1$ and Scheme $2$, hydrogen plasma is used, $i.e.$, $Z =1$ and $m_{i} = 1836m_{e}$ for ions. The temperatures of electrons and ions are $T_{e} = 300 $ eV and $T_{i} = 6 $ eV , which gives $ZT_{e}/T_{i}=50$. Compared with the plasma temperatures in recent works\cite{SBS9,exp}, the ion temperature in this paper is lower, we choose lower ion temperature to reduce the influence of Landau damping\cite{SBS4}. The plasma density ranges from $0$ to $0.04 n_{c}$.  The wavelength of SRS pump laser and SC-SBS pump laser are $\lambda_{0}=800$nm and $\lambda_{01}=1000$nm, respectively, and the wavelength of seed laser also equals to $1000$nm. Both the intensities of SRS pump laser and SC-SBS pump laser are $4\times10^{14}$ ${\rm W/cm^{2}}$, and the intensity of seed laser is $1\times10^{14}$ ${\rm W/cm^{2}}$. The initial seed laser has a Gaussian waveform with $\tau_{\rm{FWHM}} = 160$fs. Under these conditions, SBS is in the strong coupling regime for $n_{e} > 0.005 n_{c}$ based on Eq.~(\ref{SC-SBS_threshold}). The growth rates of SRS and SC-SBS are $\Gamma_{\rm{SRS}} = 0.0029 \omega_{00}$ and $\Gamma_{\rm{SC-SBS}} = 0.0011 \omega_{00}$ when $n_{e} = 0.04 n_{c}$, the growth rate of SC-SBS decreases when the plasma density gets lower.

\section{numerical solution of five-wave amplification model}\label{numerical model}

   In order to investigate the difference between these two amplification schemes, Eq.~(\ref{wkb1}) is numerically solved by Lax-Wendroff method\cite{chen1}.
   The total length of simulation box is $\rm{L}_{\rm{total}} = 1.2$ mm, the spatial and time space are discretized by $dx = 0.2 c/\omega_{00}$ and $dt = 0.2 \omega_{00}^{-1}$, respectively. The total simulation time is $8$ps, and the seed laser starts to enter the simulation box at around $4$ ps. In Scheme $2$ around $18\%$ of plasma length is used for SRS amplification.

   Fig.~\ref{two_case_2}(a) shows the the amplification process by Scheme $1$. In the linear stage, the pulse width of seed laser increases, because the maximum of seed laser moves with the half speed of light\cite{malkin1,SRS13,SBS6,SBS13,SBS8}.\begin{figure*}
    \begin{minipage}[t]{0.49\linewidth}
    \centering
    \includegraphics[width=3.5in]{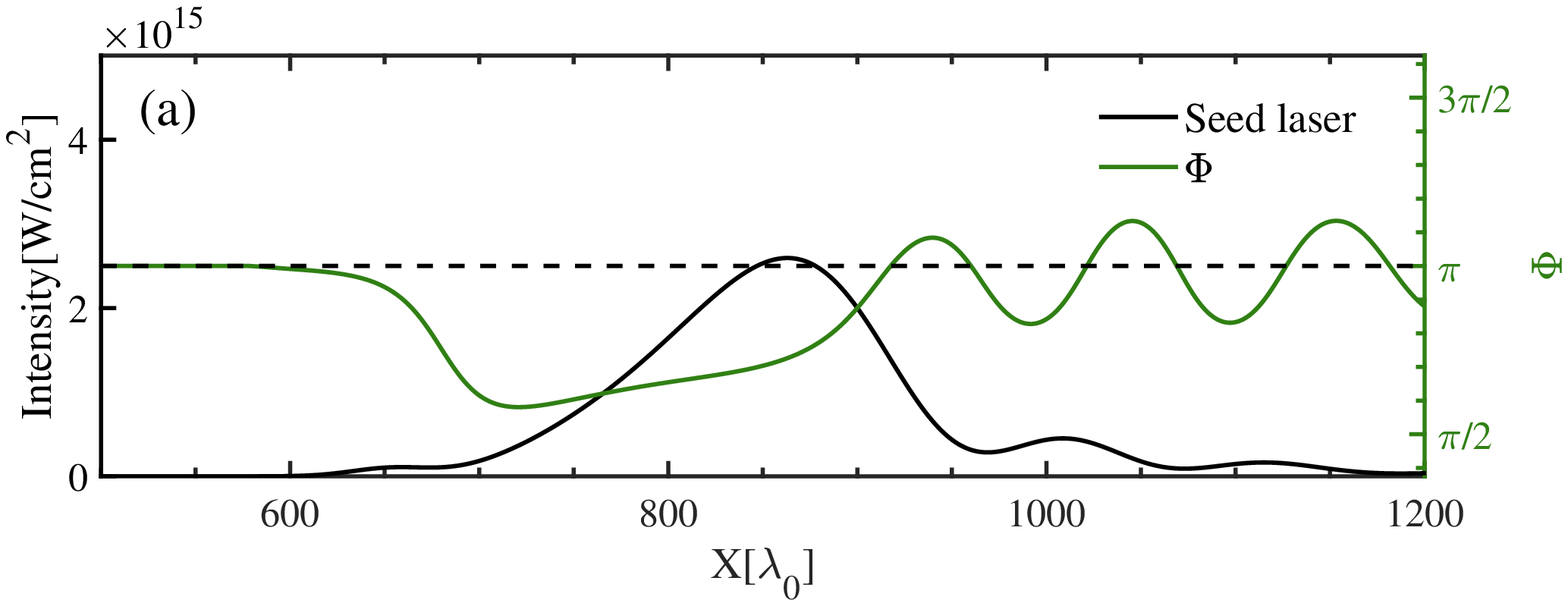}\vspace{-20pt}
    \end{minipage}%
    \begin{minipage}[t]{0.49\linewidth}
    \centering
    \includegraphics[width=3.5in]{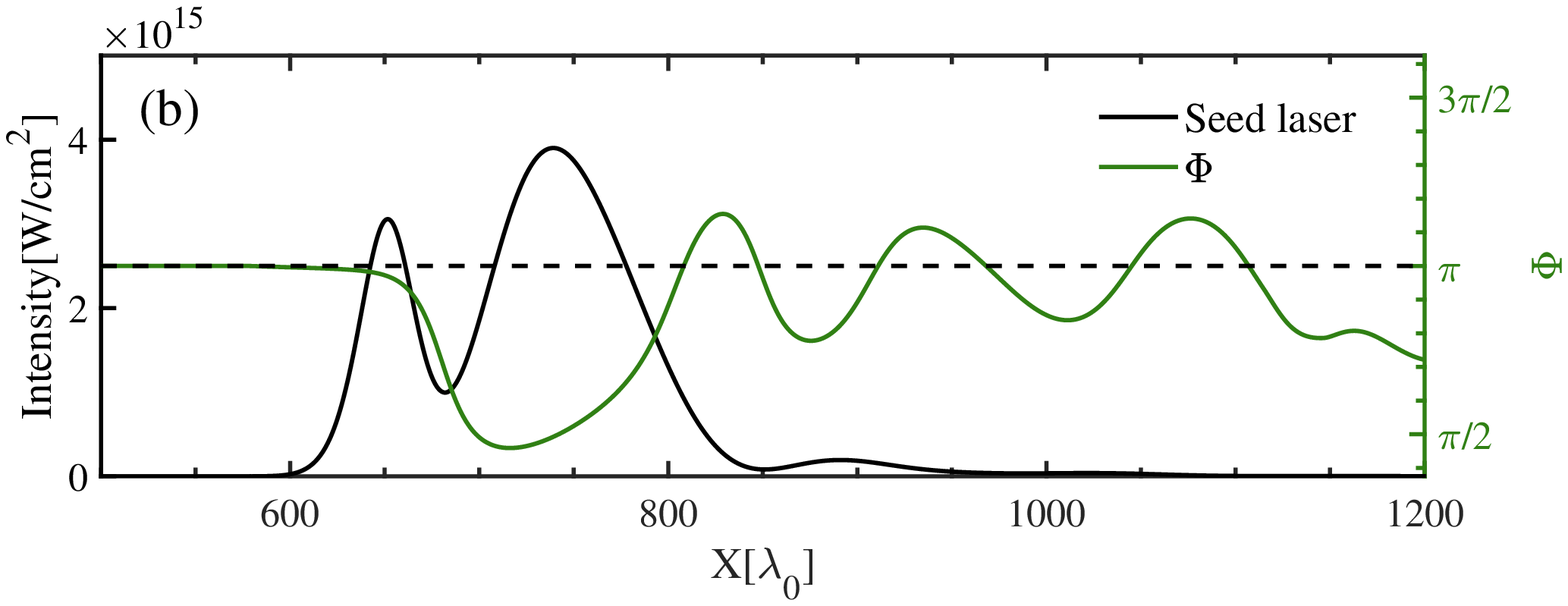}\vspace{-20pt}
    \end{minipage}
    \caption{\label{two_case_phase} Numerical solution of Eq.~(\ref{wkb1}), (a) results of Scheme $1$, black line is the seed laser at $t = 6.5$ ps, green line is the corresponding total phase in SC-SBS amplification. (b) results of Scheme $2$, black line is the seed laser at $t = 6.5$ ps, green line is the corresponding total phase in SC-SBS amplification.}
\end{figure*}Since the initial pulse width of seed laser is much smaller than that of effective amplification width\cite{SRS13,SBS13}, $\tau_{\rm{eff}} = 1/\Gamma_{\rm{SCSBS}}$, the second pulse is formed and the intensity of the first peak still equals to $1\times10^{14}$ ${\rm W/cm^{2}}$ shown by the green area in Fig.~\ref{two_case_2}(a). The duration of SBS peak is $\tau_{a} = 1076$fs (duration between two dashed lines) at $t=5.7$ ps. The duration of SBS peak reduces in the nonlinear stage, when the pump depletion happens\cite{SRS2,SRS3}. At $ t =8.0$ ps, the duration of SBS peak is $675$ fs and the intensity of seed laser increases to $6.02\times10^{15}$ ${\rm W/cm^{2}}$.

   The results of Scheme $2$ are shown in Fig.~\ref{two_case_2} (b). At $t = 5.0$ ps, the seed laser has just entered the SC-SBS region, and it has been amplified to around $3.0\times10^{15}$ ${\rm W/cm^{2}}$ by SRS.   The $\pi$ pulse of SRS is also observed due to the pump depletion, and the intensity of $\pi$ pulse is $3.2\times10^{14}$ ${\rm W/cm^{2}}$. When pump depletion happens, the spatial derivative in seed laser equation can be neglected, then the equation reduces to sine-Gordon equation, Malkin $et.al.$ found the particular solution of this equation that the seed laser repeats itself as time, which is called self-similar solution, or $\pi$-pulse solution\cite{malkin1,malkin2}.

    We find a fact that the structure of seed laser after SRS amplification is similar to the green area in Fig.~\ref{two_case_2}(a); the leading peak of seed laser barely changes in the SC-SBS region, but the second peak is responsible for the amplification. We believe that the SBS peaks are seeded by the leading peak of SRS, because SBS peak appears after the delay time it takes to develop.

   Another finding in Fig.~\ref{two_case_2} (b) is that the pulse duration is shorter, $\tau_{b} = 488$fs (duration between two dashed lines), and the intensity of seed laser grows faster than Scheme $1$. At $t = 8.0$ ps, seed laser is amplified to $8.17\times10^{15}$ ${\rm W/cm^{2}}$, which is $36\%$ higher than the intensity in Fig.~\ref{two_case_2}(a), the duration of SBS peak is $408$ fs, which is shorter than that in Scheme $1$. Next, we will discuss reasons why these phenomena happen.

   Chiaramello $et$ $al.$ studied the SC-SBS by the phase analysis of three waves\cite{Amir,SBS10}. We decide to follow their method by separating the amplitudes and phases of three waves in SC-SBS, the amplitude and phase equations can be written as,
   \begin{equation} \label{phase_equation}
   \begin{split}
      & (\partial_{t}+V_{01}\partial_{x})|a_{01}|=-\mu |a_{1}||\delta n_{iaw}|\rm{sin}(\Phi),\\
       &(\partial_{t}+V_{01}\partial_{x})\varphi_{01}=-\mu |\delta n_{iaw}|\frac{|a_{1}|}{|a_{01}|}\rm{cos}(\Phi),\\
       &(\partial_{t}+V_{1}\partial_{x})|a_{1}|=\mu |a_{01}||\delta n_{iaw}|\rm{sin}(\Phi),\\
       &(\partial_{t}+V_{1}\partial_{x})\varphi_{1}=-\mu |\delta n_{iaw}|\frac{|a_{01}|}{|a_{1}|}\rm{cos}(\Phi),\\
       &\partial_{t}^{2}|\delta n_{iaw}|-|\delta n_{iaw}|(\partial_{t}\varphi)^{2}-V_{3}^{2}(\partial_{x}^{2}|\delta n_{iaw}|\\
       &-|\delta  n_{iaw}|(\partial_{x}\varphi)^{2})=-\Lambda |a_{01}||a_{1}|\rm{cos}(\Phi),\\
       &|\delta  n_{iaw}|\partial_{t}^{2}\varphi+2\partial_{t}|\delta  n_{iaw}|\partial_{t}\varphi-V_{3}^{2}(2\partial_{x}|\delta  n_{iaw}|\partial_{x}\varphi\\
       &+|\delta  n_{iaw}|\partial_{x}^{2}\varphi)=-\Lambda |a_{01}||a_{1}|\rm{sin}(\Phi),\\
   \end{split}
   \end{equation}
   where $\mu = \frac{\omega_{00}}{4\omega_{1}}$ and $\Lambda = \frac{Z\beta n_{e}(x)k_{iaw}^{2}c^{2}}{2 n_{c}\omega_{00}^{2}}$. The amplitudes of three waves, $|a_{01}|$,  $|a_{1}|$ and $|\delta n_{iaw}|$ in Eq.~(\ref{phase_equation}), are real and positive, $\varphi_{01}$, $\varphi_{1}$ and $\varphi$ are corresponding phases, $\Phi$ is the total phase, which is obtained by $\Phi=\rm{mod}(\varphi_{01}-\varphi_{1}-\varphi,2\pi)$.

   From Eq.~(\ref{phase_equation}), we know that the energy flows from pump laser to seed laser when $\rm{sin}(\Phi)>0$, leading to the seed laser growing in the exponential stage\cite{Amir}, while the energy flows from seed laser to pump laser when $\rm{sin}(\Phi)<0$, which appears when the pump laser is depleted, and the $\pi$ pulses appear.

   The initial condition of phases are $\varphi_{01}(t_{0}) = 0$, $\varphi_{1}(t_{0}) = 0$  and $\varphi(t_{0}) = -\pi$, so $\Phi(t_{0}) = \pi$. At the early stage when the pump laser just collides with the seed laser, the phases of pump laser and ion acoustic wave keep constant, only $\varphi_{1}$ changes, thus $\Phi(t)=\pi-\varphi_{1}(t)$. When the pump depletion occurs, the phase change of pump laser should be considered, and the total phase $\Phi$ will change back to $\pi$. Once $\Phi$ becomes larger than $\pi$, the energy will flows from seed laser back to pump laser, $\pi$ pulses of seed laser appear. The length of the first interval of the total phase corresponds to the pulse width of seed laser.  In Fig.~\ref{two_case_phase}, we obtain the total phase of Scheme $1$ and Scheme $2$ at $t=6.5$ps. We observe that the width of the first interval in Scheme 2 is shorter than Scheme 1. Based on       Amiranoff's work\cite{Amir}, the total phase changes faster when the initial intensity of seed laser is higher, and the pulse width of seed laser is related to $I_{\rm{s0}}/I_{\rm{p}}$\cite{Amir},
    \begin{equation} \label{pulse_width}
     l_{interval} = |V_{1}|\tau_{\rm{eff}}\frac{6}{ I_{\rm{s0}}/I_{\rm{p}}+5 },
   \end{equation}where $I_{\rm{s0}}$ is the intensity of seed laser when it enter into the SC-SBS region, and $I_{\rm{p}}$ is the intensity of pump laser. In scheme $2$, $I_{\rm{s0}}/I_{\rm{p}} = 7.5$, so the pulse width in Scheme 2 is nearly half of that in Scheme 1.

   We also obtain that the energy transfer efficiency, defined by $\eta = \int_{0}^{\rm{\tau_{s}}}I_{\rm{seed}}(t)dt/\int_{0}^{\rm{t_{total}}} I_{\rm{pump}}(t)dt$ , where $\tau_{s}$ is the duration of seed laser at the end of simulation, for these two schemes in Fig.~\ref{two_case_2} are both around $70\%$. Therefore, it is easy to understand the higher intensity of seed laser obtained in Scheme 2, since its pulse width in Scheme 2 is shorter. Another evidence is that in Fig.~\ref{two_case_phase}(b), the minimum value of $\Phi$ in Scheme $2$ is closer to $\pi/2$, or $\rm{sin}(\Phi)$ is closer to $1$, which explains why the seed laser grows faster than Scheme $1$.

\begin{figure*}
    \begin{minipage}[t]{0.487\linewidth}
    \centering
    \includegraphics[width=3.65in]{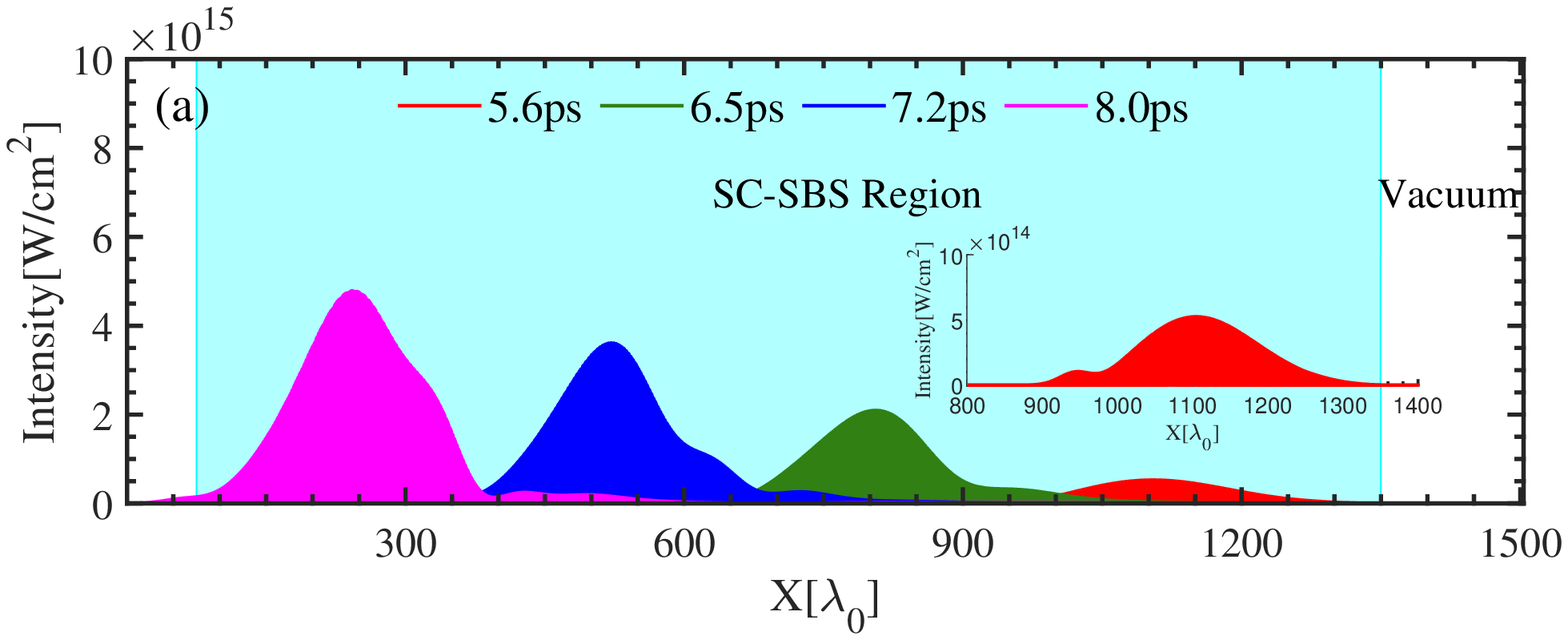}\vspace{-20pt}
    \end{minipage}%
    \begin{minipage}[t]{0.487\linewidth}
    \centering
    \includegraphics[width=3.65in]{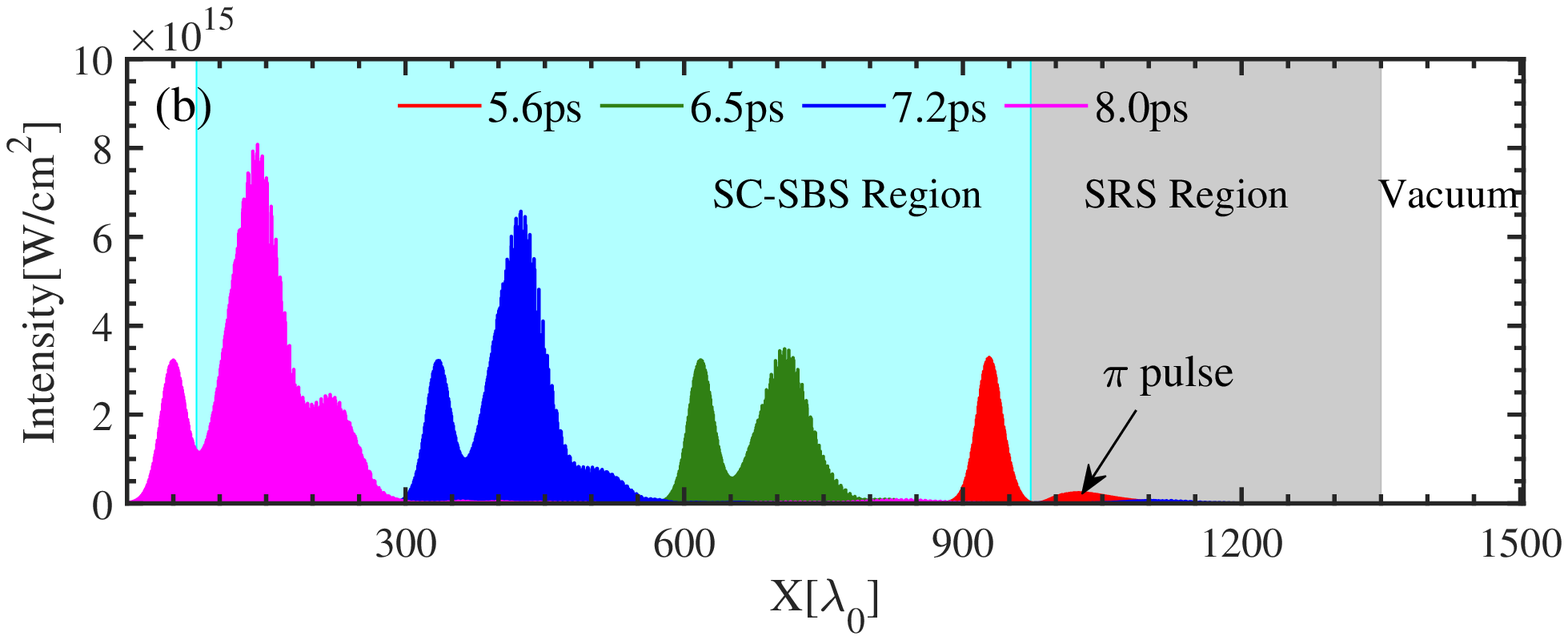}\vspace{-20pt}
    \end{minipage}

    \caption{\label{vlasov_amp} Vlama simulation results. (a) is the amplification results of Scheme $1$, inset figure is the seed laser at $t = 5.6$ ps. (b) is the amplification results of Scheme $2$ with $\rm{L}_{\rm{SRS}}/\rm{L}_{\rm{total}}=0.25$. }
\end{figure*}

   Besides, we also observe that the slope of the second peak in Scheme $2$ is steeper than that in Scheme $1$. In the early work about SRS amplification, Tsidulko $et$ $al.$ found that the intensity and shape of the final amplified pulses are closely related to the local slope of the wave front of seed laser\cite{tsi}; the steeper the slope of wave front results in better amplification. In the SRS region of Scheme 2, the slope of seed front becomes steeper because of seed compression. \begin{figure}[htbp]
    \begin{center}
      \includegraphics[width=0.47\textwidth,clip,angle=0]{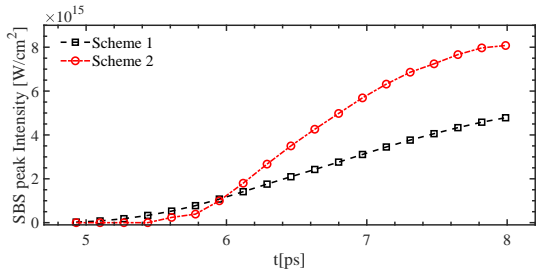}\vspace{-5pt}
      \caption{\label{seed_time} Time dependence of SBS peak intensity in Vlama simulations, the black square stands for  Scheme  $1$ and the red circle represents the results of  Scheme $2$.}
    \end{center}
  \end{figure}In Fig.~\ref{two_case_2}(b), the slope of the wave front of SBS peak is nearly equal to that of first peak $i.e.$ SRS peak. Thus, in the Scheme $2$ the SRS amplification  not only improves the intensity, but also steepens the seed front, which are both favorable to SC-SBS amplification.

  We have considered to change the ordering of the pump pulses and adapt correspondingly, the seed amplification is no better than that in scheme 2. Under the plasma conditions and laser conditions in this paper, Raman amplification first then SC-SBS amplification is the better choice.

  The precursor in scheme 2 contains part of energy, which is a disadvantage, however, the SBS peak is seeded by the leading peak of SRS, higher intensity of precursor results in faster growth of SBS peak. Thus, the precursor has a double-edged effect on SC-SBS amplification.

   The envelope coupling model in Eq.~(\ref{wkb1}) is a simplified model to analyse the SRS and SC-SBS amplification since it does not consider some nonlinear effects, such as harmonic waves, particle trapping, and nonlinear Landau damping. In the next two sections, we will use fully kinetic simulation codes, Vlasov-Maxwell code (Vlama) and PIC code(EPOCH), to verify the differences of these two schemes.


\section{one dimensional kinetic Vlasov-Maxwell simulations }\label{Simulation model}

Plasma parameters in the Vlama simulations are the same to that in Sec.~\ref{numerical model}.\begin{figure}[htbp]
    \begin{center}
      \includegraphics[width=0.47\textwidth,clip,angle=0]{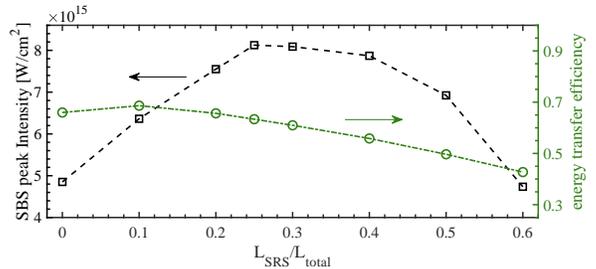}\vspace{-5pt}
      \caption{\label{srs_scan} Vlama simulation results, Black square are the SBS peak intensities varies with $\rm{L}_{\rm{SRS}}/\rm{L}_{\rm{total}}$.Green circles are the energy transfer efficiency varies with $\rm{L}_{\rm{SRS}}/\rm{L}_{\rm{total}}$. }
    \end{center}
  \end{figure}   $5\%$ vacuum space is reserved on the left of the simulation box and  $10\%$ vacuum space on the right, $\rm{L}_{\rm{SRS}}/\rm{L}_{\rm{total}} = 0.25$, where $\rm{L}_{\rm{total}} = 1.2$ mm. The conditions for lasers are also the same to that in Sec.~\ref{numerical model}. The space and time are discretized by $dx = 0.2c/\omega_{0}$ and $dt = 0.2 \omega_{0}^{-1}$. The velocity space of electrons is $[-0.5c,0.5c]$, and the velocity space of ions is $[-0.002c,0.002c]$. The space-velocity mesh grid of electrons and ions distribution functions are both $47232\times2049$.

  \begin{figure*}
    \begin{minipage}[t]{0.48\linewidth}
    \centering
    \includegraphics[width=3.5in]{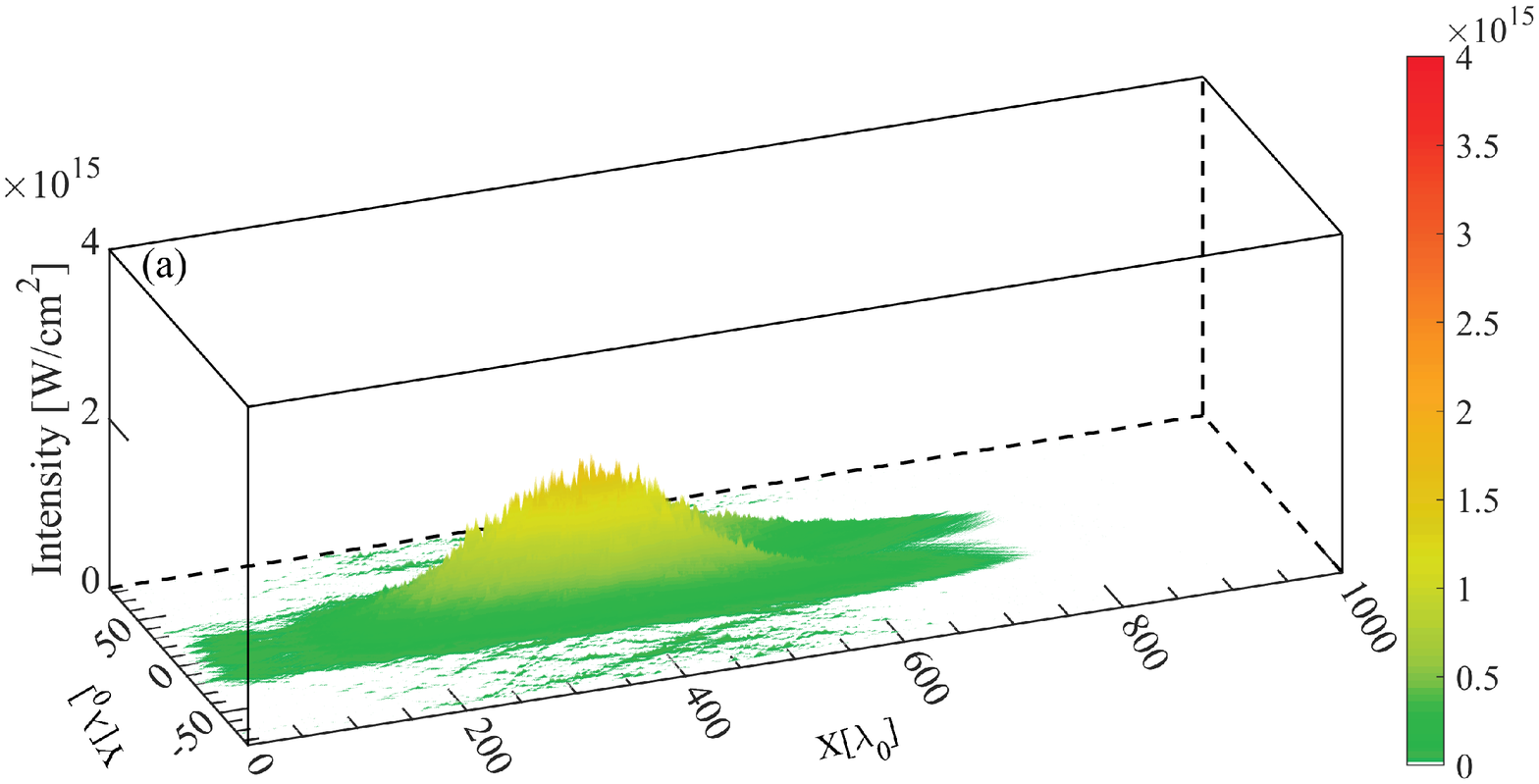}\vspace{-20pt}
    \end{minipage}%
    \begin{minipage}[t]{0.48\linewidth}
    \centering
    \includegraphics[width=3.5in]{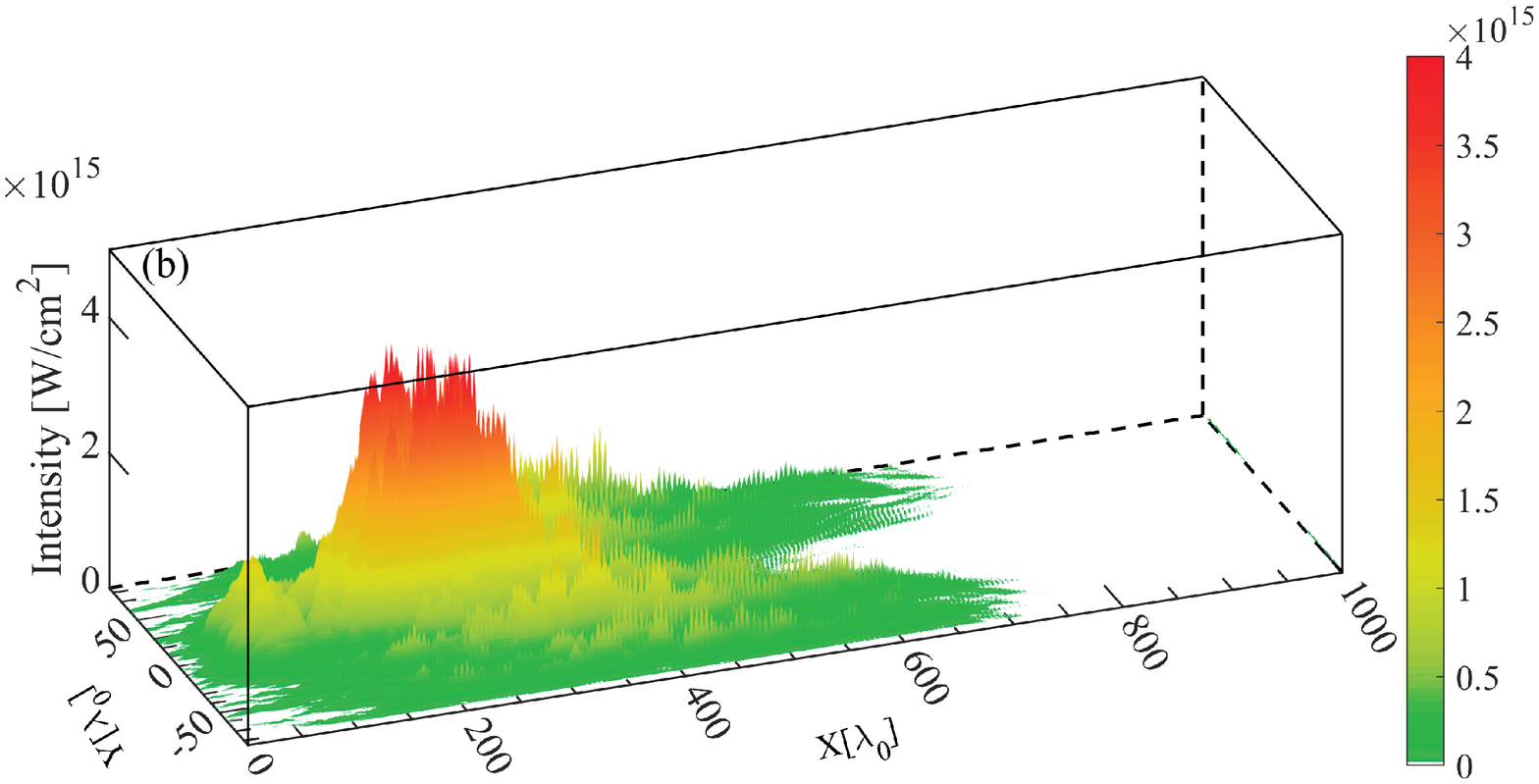}\vspace{-20pt}
    \end{minipage}

    \caption{\label{pic_2d} 2D PIC simulation results, (a) is the amplified seed laser in pure SC-SBS scheme,(b)is the amplified seed laser in new scheme. }
\end{figure*}

Fig.~\ref{vlasov_amp} (a) is the amplification process by Scheme $1$, similar to the results of five-wave simulations. At the linear stage, the duration of seed laser is widened, and there are two peaks of seed laser  formed due to shorter initial seed duration than $1/\Gamma_{\rm(SC-SBS)}$ shown in the red line at $t = 5.6$ ps. Then, the duration of seed laser has a little change at the nonlinear stage. After amplification, the maximum intensity of seed laser becomes $4.8\times10^{15}$ ${\rm W/cm^{2}}$. Fig.~\ref{vlasov_amp}(b) shows the results of Scheme $2$. After SRS amplification, the intensity of leading peak becomes $3.3\times10^{15}$ ${\rm W/cm^{2}}$, and the $\pi$ pulse of SRS is also observed. The intensity of $\pi$ pulse is $2.4\times10^{14}$ ${\rm W/cm^{2}}$. The leading peak of SRS will seed SC-SBS amplification in the next stage. In the SC-SBS region, the intensity of the leading peak does not change the second peak is amplified, which agrees with our five-wave simulations in Fig.~\ref{two_case_2}. The pulse duration of seed in Fig.~\ref{vlasov_amp}(b) is shorter than that in Fig.~\ref{vlasov_amp}(a), and the maximum intensity of the new scheme is $8.12\times10^{15}$ ${\rm W/cm^{2}}$, which is much higher than that in Scheme $1$. The time dependence of SBS peak intensities of two schemes are collected and shown in Fig.~\ref{seed_time}. The growth of seed laser in Scheme $2$ is faster than that in Scheme $1$ in both linear stage (exponential growth) and nonlinear stage (linear growth). This is because the length reduction of first phase interval discussed in Sec.~\ref{numerical model}.

Besides, by changing the ratio of SRS length, $i.e.$ $\rm{L}_{\rm{SRS}}/\rm{L}_{\rm{total}}$, we find that the SBS peak intensity increases first then decreases with $\rm{L}_{\rm{SRS}}/\rm{L}_{\rm{total}}$ as shown in the black squares of Fig.~\ref{srs_scan}. The maximum intensity of SBS peak is $8.12\times10^{15}$ ${\rm W/cm^{2}}$ when $\rm{L}_{\rm{SRS}}/\rm{L}_{\rm{total}} = 0.25$. However, the SBS peak intensity decreases when $\rm{L}_{\rm{SRS}}/\rm{L}_{\rm{total}}> 0.25$, because the SC-SBS amplification length reduces with the increase of $\rm{L}_{\rm{SRS}}/\rm{L}_{\rm{total}}$, leading to the existence of the maximum SBS peak intensity in Fig.~\ref{srs_scan}.

As shown by the green circles in Fig.~\ref{srs_scan}, the energy transfer efficiency $\eta$ in Vlama simulations are also obtained. Normally, $\eta$ in SC-SBS is higher than that in SRS \cite{SBS2,SBS4}, because based on Manley-Rowe relations, the frequency of ion acoustic wave is much less than the frequency of pump laser in SC-SBS,  so the energy loss to plasma wave is lower than SRS.  The energy transfer efficiency of pure SC-SBS scheme is around $66\%$, and it decreases with the $\rm{L}_{\rm{SRS}}/\rm{L}_{\rm{total}}$.  We also observe that when $\rm{L}_{\rm{SRS}}/\rm{L}_{\rm{total}} =0.25$, $ \eta = 63\%$ , which  only decreases $3\%$, compared with Scheme $1$. Thus, in Scheme $2$ the maximum intensity of SBS peak increases, but the energy transfer efficiency is not much lower than that of SC-SBS. The results of Vlama simulations  agree with the five-wave amplification model very well.

\section{two dimensional PIC simulations }\label{PIC model}

In order to verify our new scheme in higher dimensions, we turn to use two dimensional and fully kinetic PIC simulation code EPOCH\cite{pic}. The plasma type and temperature are the same with that in Vlasov simulations. The simulation box is $\rm{L}_{x} = 0.8$mm, and $\rm{L}_{y} = 128$ $\rm{\mu m}$, and there are $30$ cells per $\lambda_{0}$ at the longitudinal direction and $8$ cells per $\lambda_{0}$ at the transverse direction. The intensities of SBS pump laser and SRS pump laser are both $4\times10^{14}$ ${\rm W/cm^{2}}$, and they are simple plane wave. The intensity of seed laser is a Gaussian laser with an intensity of $1\times10^{14}$ ${\rm W/cm^{2}}$, $\rm{FWHM} = 160$fs, and the pulse width in transverse dimension of $16$${\rm {\mu m}}$. $5\%$ vacuum space on both sides of the simulation box is remained.  In the pure SC-SBS amplification case, the plasma density ranges from $0$ to $0.04 n_{c}$, and there is $20\%$ uniform density for SRS amplification in Scheme 2. The particle number per cell (PPC) for electrons and ions are both $100$.

Fig.~\ref{pic_2d} (a) illustrates the results of 2D PIC simulations for Scheme $1$. The duration of seed laser is larger than that of Scheme 2 shown in Fig.~\ref{pic_2d} (b). After interaction, the intensity of seed laser peak intensity is $1.74\times10^{15}$ ${\rm W/cm^{2}}$ and the maximum intensity in  Scheme 2 is $4.28\times10^{15}$ ${\rm W/cm^{2}}$. The maximum intensity is more than double with the pure SC-SBS scheme. In Fig.~\ref{pic_2d} (b), the first peak of seed laser, $i.e.$ SRS peak is clearly observed, and its intensity has a little change through the SC-SBS amplification, which agrees well with the previous analyses. However, the filamentation of seed laser is observed in Fig.~\ref{pic_2d} (b). Filamentation effects are also observed in the early works\cite{SBS3,SBS11,si}, because of the ponderomotive effect, or the thermal effect, which makes the refractive index toward the center at the transverse direction.
We obtain the theoretical growth rate of filamentation for seed laser by $\gamma_{filam}=1/8(v_{osc}/v_{e})^{2}\omega_{pe}^{2}/\omega_{0}=0.025\omega_{0}$ (when the intensity of seed laser is $4.0\times10^{15}\rm{W/cm^{2}}$)\cite{kruer}. In Fig.~\ref{pic_2d} (b), the filamentation growth rate for simulation is $\gamma_{filam}=0.0242\omega_{0}$. Thus, the filamentation is mostly related to the high intensity of seed laser. Though with the filamentation, Scheme $2$ still has a better performance than Scheme $1$. In the future work, we will focus on the filamentation effects in Scheme $2$.

\section{Conclusion and discussion}\label{conclusion}

 In this paper, at first, we propose a scheme to improve SC-SBS amplification by the assisted of SRS amplification, and construct a five-wave amplification model, then we numerically solve the five-wave envelope equations to study the influence of SRS amplification on SC-SBS amplification. Comparing with pure SC-SBS amplification, our scheme increases the maximum intensity of seed laser and reduces the pulse width of seed laser, and the $\pi$ pulse of SRS can seed the SC-SBS amplification. We find that the reason for this phenomenon is that (i) the width of the first interval of the total phase becomes shorter in the new scheme because of the higher intensity of the seed laser when it enters the SC-SBS region; (ii) the energy transfer efficiency changes a little compared with pure SC-SBS scheme. Next, the Vlama simulations are carried out to test our schemes. The maximum intensity of seed laser is improved nearly $69\%$ in the new scheme, the optimal $\rm{L}_{\rm{SRS}}/\rm{L}_{\rm{total}}$ is tested as $0.25$, and the energy transfer efficiency changes little when $\rm{L}_{\rm{SRS}}/\rm{L}_{\rm{total}}=0.25$ compared with the pure SC-SBS amplification. At last, the two dimensional PIC code is used to verify our conclusion, which qualitatively agrees with our predications.

 As is well known, SC-SBS has higher energy transfer efficiency, but it has wider pulse width than SRS\cite{SBS5,SBS6,SBS8,SBS10}. Our scheme reduces the width of first interval of total phase, so the intensity of seed laser is easier to increase at the nonlinear stage than pure SC-SBS scheme. The SC-SBS pump laser, SRS pump laser, and seed laser in our scheme are easy to get in experiments, because their wavelengthes are close to the wavelengthes of common laser devices. Besides, in experiments, the plasma profiles are usually Gaussian type\cite{exp}, the plasma density change rate is small at the center, it is possible to implement SRS amplification at this part of plasma. The new scheme does not require extra input energy from the pump laser, also the initial seed laser does not require additional processing.

\section*{Acknowledgements}
We are pleased to acknowledge useful discussions with Y. G. Chen, S. X. Xie, N. Peng, W. B. Yao and S. Tan.  This work was supported by the Strategic Priority Research Program of Chinese Academy of Sciences (Grant No.XDA25050700), National Natural Science Foundation of China (Grant Nos.11805062, 11875091 and 11975059), Science Challenge Project, No. TZ2016005 and Natural Science Foundation of Hunan Province, China (Grant No.2020JJ5029).


\begin{thebibliography}{0}%
\makeatletter
\providecommand \@ifxundefined [1]{%
 \@ifx{#1\undefined}
}%
\providecommand \@ifnum [1]{%
 \ifnum #1\expandafter \@firstoftwo
 \else \expandafter \@secondoftwo
 \fi
}%
\providecommand \@ifx [1]{%
 \ifx #1\expandafter \@firstoftwo
 \else \expandafter \@secondoftwo
 \fi
}%
\providecommand \natexlab [1]{#1}%
\providecommand \enquote  [1]{``#1''}%
\providecommand \bibnamefont  [1]{#1}%
\providecommand \bibfnamefont [1]{#1}%
\providecommand \citenamefont [1]{#1}%
\providecommand \href@noop [0]{\@secondoftwo}%
\providecommand \href [0]{\begingroup \@sanitize@url \@href}%
\providecommand \@href[1]{\@@startlink{#1}\@@href}%
\providecommand \@@href[1]{\endgroup#1\@@endlink}%
\providecommand \@sanitize@url [0]{\catcode `\\12\catcode `\$12\catcode
  `\&12\catcode `\#12\catcode `\^12\catcode `\_12\catcode `\%12\relax}%
\providecommand \@@startlink[1]{}%
\providecommand \@@endlink[0]{}%
\providecommand \url  [0]{\begingroup\@sanitize@url \@url }%
\providecommand \@url [1]{\endgroup\@href {#1}{\urlprefix }}%
\providecommand \urlprefix  [0]{URL }%
\providecommand \Eprint [0]{\href }%
\providecommand \doibase [0]{http://dx.doi.org/}%
\providecommand \selectlanguage [0]{\@gobble}%
\providecommand \bibinfo  [0]{\@secondoftwo}%
\providecommand \bibfield  [0]{\@secondoftwo}%
\providecommand \translation [1]{[#1]}%
\providecommand \BibitemOpen [0]{}%
\providecommand \bibitemStop [0]{}%
\providecommand \bibitemNoStop [0]{.\EOS\space}%
\providecommand \EOS [0]{\spacefactor3000\relax}%
\providecommand \BibitemShut  [1]{\csname bibitem#1\endcsname}%
\let\auto@bib@innerbib\@empty
\end{thebibliography}%


\begin{thebibliography}{100}
\newcommand{\DOI}[1]{doi: \href{https://doi.org/#1}{#1}}

\bibitem{malkin1}V. M. Malkin, G. Shvets and  N. J. Fisch,  Phys. Rev. Lett. {\bf 82}, 4448 (1999). (\DOI{10.1103/PhysRevLett.82.4448})
\bibitem{malkin2}V. M. Malkin, G. Shvets and  N. J. Fisch,  Phys. Rev. Lett. {\bf 84}, 1208 (2000).(\DOI{10.1103/PhysRevLett.84.1208})
\bibitem{cpa} D. Strickland and  G. Mourou, Opt. Commun. {\bf 55}, 447 (1985).(\DOI{10.1016/0030-4018(85)90151-8})
\bibitem{tsi}Yu. A. Tsidulko, V. M. Malkin and N. J. Fisch, Phys. Rev. Lett. {\bf 88}, 235004 (2002). (\DOI{10.1103/PhysRevLett.88.235004})

\bibitem{2dSRS}G. M. Fraiman, N. A. Yampolsky, V. M. Malkin and N. J. Fisch,  Phys. Plasmas {\bf 9}, 3617 (2002).(\DOI{10.1063/1.1491959})

\bibitem{SRS1} R. M. G. M. Trines, F. Fi$\acute{u}$za, R. Bingham, R. A. Fonseca, L. O. Silva, R. A. Cairns and P. A. Norreys, Nature Phys. {\bf 7}, 87 (2011).(\DOI{10.1038/nphys1793})
\bibitem{SRS2} J. Ren, W. Cheng, S. Li and S. Suckewer, Nature Phys. {\bf 3}, 732 (2007).(\DOI{10.1038/nphys717})
\bibitem{SRS3} Z. Toroker, V.M. Malkin and  N. J. Fisch, Phys. Rev. Lett. {\bf 109}, 085003 (2012).(\DOI{10.1103/PhysRevLett.109.085003})
\bibitem{SRS4} B. Ersfeld and  D. A. Jaroszynski, Phys. Rev. Lett. {\bf 95}, 165002 (2005).(\DOI{10.1103/PhysRevLett.95.165002})
\bibitem{SRS5} R. Nuter and  V. Tikhonchuk, Phys. Rev. E {\bf 87}, 043109 (2013).(\DOI{10.1103/PhysRevE.87.043109})
\bibitem{SRS6} K. Qu, I. Barth and  N. J. Fisch, Phys. Rev. Lett. {\bf 118}, 164801 (2017).(\DOI{10.1103/PhysRevLett.118.164801})
\bibitem{SRS7} I. Barth and N. J. Fisch, Phys. Rev. E. {\bf 97}, 033201 (2018).(\DOI{10.1103/PhysRevE.97.033201})
\bibitem{SRS8} M. R. Edwards, Y. Shi, J. M. Mikhailova and  N. J. Fisch, Phys. Rev. Lett. {\bf 123}, 025001 (2019).(\DOI{10.1103/PhysRevLett.123.025001})
\bibitem{SRS9} M. R. Edwards, K. Qu, J. M. Mikhailova and  N. J. Fisch, Phys. Plasmas {\bf 24}, 103110 (2017).(\DOI{10.1063/1.4997246})
\bibitem{SRS10}Z. Wu, Q. Chen, A. Morozov and S. Suckewer, Phys. Plasmas {\bf 26}, 103111 (2019).(\DOI{10.1063/1.5094744})
\bibitem{SRS11}D. Haberberger, A. Davies, J. L. Shaw, R. K. Follett, J. P. Palastro and  D. H. Froula, Phys. Plasmas {\bf 28}, 062311 (2021).(\DOI{10.1063/5.0049222})
\bibitem{SRS12}Y. G. Chen, Y. Chen, S. X. Xie, N. Peng, J. Q. Yu and  C Z Xiao,  Chinese Physics B {\bf 30} 105202 (2021).(\DOI{10.1088/1674-1056/ac0521})
\bibitem{SRS13}G. Lehmann, K. H. Spatschek and  G. Sewell, Phys. Rev. E {\bf 87},063107 (2013).(\DOI{10.1103/PhysRevE.87.063107})
\bibitem{SRS14}Y. Ping, W. Cheng, S. Suckewer, D. S. Clark and  N. J. Fisch, Phys. Rev. Lett. {\bf 92}, 175007 (2004).(\DOI{10.1103/PhysRevLett.92.175007})
\bibitem{SRS15}C.-H. Pai, M.-W. Lin, L.-C. Ha, S.-T. Huang, Y.-C. Tsou, H.-H. Chu, J.-Y. Lin, J. Wang and S.-Y. Chen, Phys. Rev. Lett. {\bf 101}, 065005 (2008).(\DOI{10.1103/PhysRevLett.101.065005})
\bibitem{SRS16} Z. Toroker, V. M. Malkin and N. J. Fisch, Phys. Plasmas {\bf 21}, 113110 (2014)(\DOI{10.1063/1.4902362})
\bibitem{SRS17} R. M. G. M. Trines, F. Fi$\acute{u}$za, R. Bingham, R. A. Fonseca, L. O. Silva, R. A. Cairns and P. A. Norreys, Phys. Rev. Lett. {\bf 107}, 105002 (2011).(\DOI{10.1103/PhysRevLett.107.105002})


\bibitem{rand} A. A. Solodov, V. M. Malkin and N. J. Fisch, Phys. Plasmas {\bf 10}, 2540 (2003).(\DOI{10.1063/1.1576761})

\bibitem{Riconda2d} C. Riconda, S. Weber, L. Lancia, J.-R. Marqu$\grave{e}$s, G. Mourou and J. Fuchs, Phys. Plasmas Control. Fusion {\bf 57},  014002(2015) (\DOI{10.1088/0741-3335/57/1/014002})

\bibitem{SBS1} M. R. Edwards, N. J. Fisch, and J. M. Mikhailova, Phys. Rev. Lett. {\bf 116}, 015004 (2016).(\DOI{10.1103/PhysRevLett.116.015004})
\bibitem{SBS2} M. R. Edwards, Q. Jia, J. M. Mikhailova and N. J. Fisch, Phys. Plasmas {\bf 23}, 083122 (2016).(\DOI{10.1063/1.4961429})
\bibitem{chen} Y. Chen, C. Y. Zheng, Z. J. Liu, L. H. Cao, Q. S. Feng and  C. Z. Xiao, Phys. Plasmas Control. Fusion {\bf 62}, 105020 (2020).(\DOI{10.1088/1361-6587/ab98df})
\bibitem{SBS3} S. Weber, C. Riconda, L. Lancia, J.-R. Marqu$\grave{e}$s, G. A. Mourou and J. Fuchs, Phys. Rev. Lett. {\bf 111}, 055004 (2013).(\DOI{10.1103/PhysRevLett.111.055004})
\bibitem{SBS4} A. A. Andreev, C. Riconda, V. T. Tikhonchuk and  S. Weber, Phys. Plasmas {\bf 13}, 053110 (2006).(\DOI{10.1063/1.2201896})
\bibitem{SBS5} L. Lancia, J.-R. Marqu$\grave{e}$s, M. Nakatsutsumi, C. Riconda, S. Weber, S. H$\ddot{u}$ller, A. Man$\check{c}$i$\acute{c}$,1 P. Antici, V. T. Tikhonchuk, A. H$\acute{e}$ron, P. Audebert and J. Fuchs, Phys. Rev. Lett. {\bf 104}, 025001 (2010).(\DOI{10.1103/PhysRevLett.104.025001})
\bibitem{SBS6} F. Schluck, G. Lehmann, C. M$\ddot{u}$ller and K. H. Spatschek, Phys. Plasmas {\bf 23}, 083105 (2016).(\DOI{10.1063/1.4960028})
\bibitem{SBS13} F. Schluck, G. Lehmann and K. H. Spatschek, Phys. Plasmas {\bf 22}, 093104 (2015)(\DOI{10.1063/1.4929859})
\bibitem{SBS7} Q. Jia, I. Barth, M. R. Edwards, J. M. Mikhailova and N. J. Fisch, Phys. Plasmas {\bf 23}, 053118 (2016).(\DOI{10.1063/1.4951027})
\bibitem{SBS8} G. Lehmann and  K. H. Spatschek, Phys. Plasmas {\bf 20}, 073112 (2013).(\DOI{10.1063/1.4816030})
\bibitem{SBS9}M. Chiaramello, C. Riconda, F. Amiranoff, J. Fuchs, M. Grech, L. Lancia, J.-R. Marqu$\grave{e}$s, T. Vinci and S. Weber, Phys. Plasmas {\bf 23}, 072103 (2016).(\DOI{10.1063/1.4955322})
\bibitem{SBS11}E. P. Alves, R. M. G. M. Trines, K. A. Humphrey, R. Bingham, R. A. Cairns, F. Fi$\acute{u}$za, R. A. Fonseca and L. O. Silva, Phys. Plasmas Control. Fusion {\bf 63},  114004 (2021).(\DOI{10.1088/1361-6587/ac2613})
\bibitem{SBS12}K. A. Humphrey, R. M. G. M. Trines, F. Fiuza, D. C. Speirs, P. Norreys, R. A. Cairns, L. O. Silva and R. Bingham, Phys. Plasmas {\bf 20}, 102114 (2013).(\DOI{10.1063/1.4825356})
\bibitem{Amir} F. Amiranoff, C. Riconda, M. Chiaramello, L. Lancia, J. R. Marqu$\grave{e}$s and  S. Weber, Phys. Plasmas {\bf 25}, 013114 (2018).(\DOI{10.1063/1.5019374})

\bibitem{Riconda} C. Riconda, S. Weber, L. Lancia, J.-R. Marqu$\grave{e}$s, G. A. Mourou and J. Fuchs, Phys. Plasmas {\bf 20}, 083115 (2013).(\DOI{10.1063/1.4818893})

\bibitem{leh1} G. Lehmann and K. H. Spatschek, Phys. Plasmas {\bf 22}, 043105 (2015).(\DOI{10.1063/1.4916958})
\bibitem{leh2} G. Lehmann and K. H. Spatschek, Phys. Plasmas {\bf 23}, 023107 (2016).(\DOI{10.1063/1.4941966})

\bibitem{shou} M. Shoucri, J.-P. Matte and F. Vidal, Phys. Plasmas {\bf 22}, 053101 (2015).(\DOI{10.1063/1.4919614})

\bibitem{kruer}W. L. Kruer, \emph{The Physics of Laser Plasma Interactions} (Westview Press, Boulder, 2003).
\bibitem{Nicholson} D. R. Nicholson, \emph{Introduction to Plasma Theory }(John Wiley \& Sons, New York, 1983).
\bibitem{compress1} Yu. A. Tsidulko, V. M. Malkin and  N. J. Fisch, Phys. Rev. Lett. {\bf 88},235004 (2002).(\DOI{10.1103/PhysRevLett.88.235004})
\bibitem{compress2} V. M. Malkin and N. J. Fisch, Phys. Plasmas {\bf 8}, 4698 (2001).(\DOI{10.1063/1.1400791})
\bibitem{compress3} Z. Toroker, V. M. Malkin, A. A. Balakin, G. M. Fraiman and N. J. Fisch, Phys. Plasmas {\bf 19}, 083110 (2012).(\DOI{10.1063/1.4745868})
\bibitem{compress4}D. Turnbull, S. Bucht, A. Davies, D. Haberberger, T. Kessler, J. L. Shaw and D. H. Froula, Phys. Rev. Lett. {\bf 120}, 024801 (2018).(\DOI{10.1103/PhysRevLett.120.024801})

\bibitem{SBS10} M. Chiaramello, F. Amiranoff, C. Riconda and S. Weber, Phys. Rev. Lett. {\bf 117}, 235003 (2016).(\DOI{10.1103/PhysRevLett.117.235003})
\bibitem{liu} Z. J. Liu, S. P. Zhu, L. H. Cao and C. Y. Zheng, Acta Phys. Sin. (Overseas Ed.) {\bf 56}, 7084 (2007).(\href{http://aps.cpsjournals.org.cn/CN/abstract/abstract12521.shtml}{Study of laser plasma interactions using Vlasov and Maxwell equations})

\bibitem{pic} T. D. Arber, K. Bennett, C. S. Brady, A. Lawrence-Douglas, M. G. Ramsay, N. J. Sircombe, P. Gillies, R. G. Evans, H. Schmitz, A. R. Bell and C. P. Ridgers, Plasma Phys. Control. Fusion {\bf 57}, 113001 (2015).(\DOI{10.1088/0741-3335/57/11/113001})

\bibitem{forslund}    D. W. Forslund, J. M. Kindel and E. L. Lindman, Phys. Fluids {\bf 18},1002 (1975).(\DOI{10.1063/1.861248})

\bibitem{chen1}Y. Chen, C. Y. Zheng, Z. J. Liu, L. H. Cao, Q. S. Feng, Y. G. Chen, Z. M. Huang and C. Z. Xiao,  Plasma Phys. Control. Fusion {\bf 63} 055004 (2021). (\DOI{10.1088/1361-6587/abea2f})

\bibitem{exp}  L. Lancia, A. Giribono, L. Vassura, M. Chiaramello, C. Riconda, S. Weber, A. Castan, A. Chatelain, A. Frank, T. Gangolf, M. N. Quinn, J. Fuchs and J.-R. Marqu$\grave{e}$s, Phys. Rev. Lett. {\bf 116}, 075001 (2016).(\DOI{10.1103/PhysRevLett.116.075001})


 \bibitem{si} L. L. Zhao, Y. L. Zuo, J. Q. Su and S. H. Yang, Phys. Plasmas {\bf 26}, 093102 (2019). (\DOI{10.1063/1.5094513})
 \bibitem{rmg} R. M. G. M. Trines, E. P. Alves, E. Webb, J. Vieira, F. Fi$\acute{u}$za, R. A. Fonseca, L. O. Silva, R. A. Cairns and R. Bingham, Sci. Rep. {\bf 10}, 19875 (2020). (\DOI{10.1038/s41598-020-76801-z})

\end{thebibliography}

\end{document}